\documentclass[conference]{IEEEtran}
\IEEEoverridecommandlockouts
\usepackage{cite}
\usepackage{amsmath,amssymb,amsfonts}
\usepackage{algorithmic}
\usepackage{graphicx}
\usepackage{textcomp}
\usepackage{xcolor}
\usepackage{amssymb}
\usepackage{balance}

\usepackage{booktabs} 
\usepackage[colorlinks=true, linkcolor=black, citecolor=black, urlcolor=black]{hyperref}

\def\BibTeX{{\rm B\kern-.05em{\sc i\kern-.025em b}\kern-.08em
    T\kern-.1667em\lower.7ex\hbox{E}\kern-.125emX}}
\begin{document}

\title{GasTrace: Detecting Sandwich Attack Malicious Accounts in Ethereum}

\author{
\IEEEauthorblockN{1\textsuperscript{st} Zekai Liu}
\IEEEauthorblockA{\textit{School of Cyberspace Security}\\
\textit{Hainan University}\\
Haikou, China \\
lyxxxx1988@gmail.com}
\and
\IEEEauthorblockN{2\textsuperscript{nd} Xiaoqi Li*\thanks{Corresponding author: Xiaoqi Li(csxqli@gmail.com)}}
\IEEEauthorblockA{\textit{School of Cyberspace Security}\\
\textit{Hainan University}\\
Haikou, China \\
csxqli@gmail.com}
\and
\IEEEauthorblockN{3\textsuperscript{rd} Hongli Peng}
\IEEEauthorblockA{\textit{School of Cyberspace Security}\\
\textit{Hainan University}\\
Haikou, China \\
penghongli2020@outlook.com}
\and
\IEEEauthorblockN{4\textsuperscript{th} Wenkai Li}
\IEEEauthorblockA{\textit{School of Cyberspace Security}\\
\textit{Hainan University}\\
Haikou, China \\
liwenkai871@gmail.com}
}

\makeatletter
\renewcommand\footnoterule{}
\makeatother

\maketitle

\begin{abstract}


The openness and transparency of Ethereum transaction data make it easy to be exploited by any entities, executing malicious attacks. The sandwich attack manipulates the Automated Market Maker (AMM) mechanism, profiting from manipulating the market price through front or after-running transactions. To identify and prevent sandwich attacks, we propose a cascade classification framework GasTrace. GasTrace analyzes various transaction features to detect malicious accounts, notably through the analysis and modeling of Gas features. In the initial classification, we utilize the Support Vector Machine (SVM) with the Radial Basis Function (RBF) kernel to generate the predicted probabilities of accounts, further constructing a detailed transaction network. Subsequently, the behavior features are captured by the Graph Attention Network (GAT) technique in the second classification. Through cascade classification, GasTrace can analyze and classify the sandwich attacks. Our experimental results demonstrate that GasTrace achieves a remarkable detection and generation capability, performing an accuracy of 96.73\% and an F1 score of 95.71\% for identifying sandwich attack accounts.


\end{abstract}

\begin{IEEEkeywords}
Sandwich attack, Ethereum, Gas, Cascade classification, Malicious accounts
\end{IEEEkeywords}

\section{Introduction}
Ethereum, as one of the largest decentralized platforms, has garnered significant market appeal for its openness and transactional transparency\cite{li2020characterizing,mao2024automated}. However, these certain qualities have inadvertently exposed sensitive information to potential exploitation by hackers, thereby enabling them to reap financial gain~\cite{li2021clue,zhang2022authros,li2023overview}. Among the various exploitative strategies is the sandwich attack, a sophisticated front-running and back-running strategy within cryptocurrency transactions, emblematic of Miner Extractable Value (MEV) \cite{carlsten2016instability}. This attack phenomenon was systematically identified and analyzed for the first time by Zhou, Qin, et al. \cite{zhou2021high} in 2021. Subsequent research efforts have delved into understanding and exploring the dynamics of sandwich attacks from diverse perspectives, contributing to a growing body of literature that investigates its mechanisms and potential mitigation strategies\cite{angeris2021note,bartoletti2022maximizing,torres2021frontrunner}.

The sandwich attack poses a significant defense challenge, primarily stemming from two inherent aspects of Ethereum. Firstly, the autonomy of miners or verifiers to prioritize transactions with higher Gas fees. Secondly, the platform's transparency and openness, enable malicious accounts (e.g., bots and arbitrage traders) to exploit profit in the pending transaction pool. 

While multiple studies have advocated for different defense mechanisms against sandwich attacks, it is not insufficient for detection. These approaches utilize a wide variety of means to achieve defense, such as cryptographic techniques\cite{zhou2021high}, partitioning sizable transactions into smaller ones\cite{zust2021analyzing}, implementing verification fees\cite{park2023conceptual}, adjusting slippage tolerance\cite{heimbach2022eliminating}, and limiting price impact alongside slippage\cite{park2023conceptual}. However, these solutions frequently impose strict requirements and limitations on the activities of accounts on the network, resulting in higher transaction costs.

In this paper, we present a cascading classification method to identify the potential malicious accounts in the sandwich attacks. Initially, we filter a set of Gas features to compile an initial feature list. Subsequently, we utilize a Support Vector Machine (SVM) with a Radial Basis Function (RBF) kernel to conduct preliminary classification on this feature list. Upon the first round of classification (i.e., R1 stage), we derive a set of predictive probabilities for account categories. These probabilities are integrated into the original feature list to enrich it further. Utilizing the predictive probabilities, we construct a network of nodes, where each node represents an account, and the features of the nodes include the new feature set available at the end of the R1 stage.
In the second round of classification (i.e., R2 stage), we examine the node network employing a Graph Attention Network (GAT) model~\cite{heitz2008cascaded}. This strategy aims to discern the behavioral discrepancies between malicious and legitimate accounts. By adopting this approach, we significantly improve the performance of detecting malicious accounts involved in sandwich attacks (see Sec. \ref{sec:evaluation}).



The main contributions of this paper are as follows:

\begin{itemize}

\item To the best of our knowledge, we propose the first framework GasTrace, which leverages a cascade classification approach that focuses on Gas features, identifying malicious accounts conducting sandwich attacks.

\item We construct the node networks via the predicted probabilities of accounts, which are generated by the SVM with the RBF technique in the first round.


\item We evaluate GasTrace on a dataset of 1,834 examples, and it achieves an accuracy of 96.73\% and an F1 score of 95.71\%.

\item We open-source GasTrace and experimental data on \url{https://doi.org/10.6084/m9.figshare.25142996}.

\end{itemize}

\section{GasTrace}

\subsection{Feature Extraction and Selection}

To gather the features of each account, we obtain several attributes with the account addresses, including transaction pairs, values, Gas, timestamps, and so on. Through filtering and integration of these attributes, we extract over 20 features for our analysis, such as average Gas consumption, short-term Gas volatility, and transaction frequency within a specified time range. These features reveal the potential behavior patterns from the transactions of the sandwich attacks.

Through correlational analysis, we have identified an optimal set of features for detecting malicious accounts involved in sandwich attacks. We gather a combination of 13 transactional features, with Gas-related attributes as the primary elements. These features can be divided into the Gas-related features and transaction statistics features. The Gas-related features include average Gas, average Gas fluctuation, maximum Gas fluctuation, short-term Gas fluctuation, and average ratio of Gas to value. The transaction statistics comprise the total number of transactions, transaction failure Rate, transaction creation rate, average transaction value, maximum transaction value, short-term max trades, instances of 10+ transactions in the short-term, and the ratio of zero-value to total transactions.

\subsection{First Round of Classification (R1)}
In the R1 stage, we employ an SVM model with an RBF as the kernel. The process contains regularization, dataset splitting, and optimal parameter selection.

As for the SVM, we use GridSearchCV\cite{rofik2024optimization} to perform a grid search with three hyperparameters (i.e., \textit{C}, \textit{gamma}, and \textit{class weight}), balancing the accuracy and generalization ability. Specifically, we maximize the prediction precision of the model during the optimization process, rather than the precision of the model. This operation can reduce the risk of misclassifying normal accounts as malicious ones, avoiding passing errors to the R2 stage. Finally, we obtain the SVM's predicted probabilities for each account category, which will be added to the original feature set as a new feature set $\mathbb{N}$.

Moreover, the cross-validation method is utilized during the R1 phase to ensure objective predictions of probabilities. Additionally, network edges are constructed based on the outcomes from R1, which are deliberately excluded from the direct R2 training process to minimize the risk of data leakage.

\subsection{Node Network Construction}

During the node network construction, accounts are considered nodes, and the new feature set $\mathbb{N}$ is used as node features. We connect the edges based on the prediction probability and Euclidean distance between nodes. Specifically, an edge is added between two nodes when their prediction probabilities and their Euclidean distance are within a specific range.

\begin{figure}[h!]
  \centering
  \includegraphics[width=\linewidth]{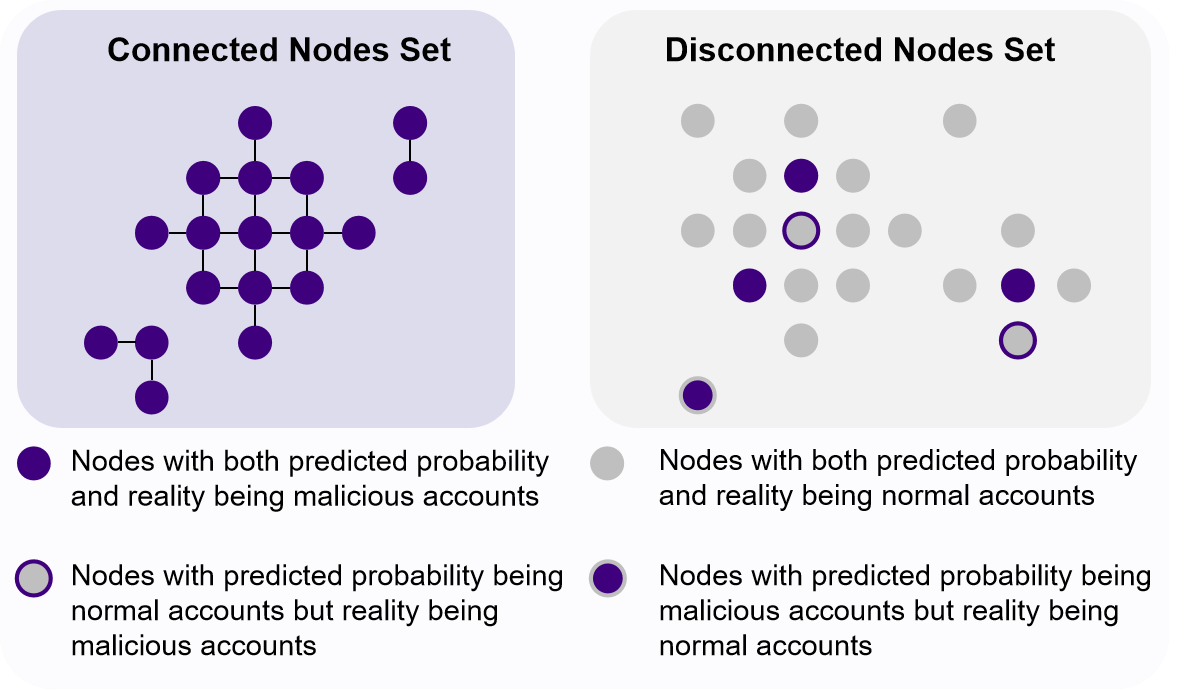}
  \caption{Example of Node Network.}
  \label{fig:node_network1}
\end{figure}

With this strategy, only a portion of the nodes predicted as malicious by R1 are connected, while other nodes appear as isolated nodes in the node network, as shown in Figure~\ref{fig:node_network1}. This method not only preserves the prediction accuracy of the R1 stage to the greatest extent but also prevents the propagation of R1 prediction errors to the subsequent R2 stage.

\subsection{Second Round of Classification (R2)}
To mitigate the limitations of the SVM, we introduce the GAT model to capture the mutual influences between accounts. 

After completing the data format conversion and dataset split, we process the new feature set $\mathbb{N}$ and adjust 6 hyperparameters for training the GAT model (i.e., the feature dimension \textit{nfeat}, the number of output classes \textit{nclass}, the dimension of the hidden layer \textit{nhid}, the learning rate \textit{lr}, the regularization methods \textit{dropout}, and \textit{weight decay}).

In this study, \textit{nfeat} and \textit{nclass} are explicitly set based on the problem definition. For the remaining 4 hyperparameters, we employ a Bayesian optimization strategy\cite{yuan2024graph} to search for their optimal values.

We devise a training function that conducts $n$ epochs of each training iteration, each epoch encompassing 5 essential steps, including resetting optimizer gradients, executing the forward pass, computing loss, performing backpropagation, and updating model parameters. During the training phase, the training function is invoked $m$ times, totaling $m \times n$ epochs. As opposed to a single execution comprising $m \times n$ iterations, this iterative approach facilitates parameter optimization after each invocation, effectively preventing overfitting.

\section{Evaluation}
\label{sec:evaluation}
\subsection{Dataset}

Our dataset comprises 464 positive examples identified as malicious account addresses involved in sandwich attacks and 1,370 negative examples as regular account addresses.

To ensure data balance during the R1 stage, 30\% of both positive and negative examples are randomly allocated to the test set, with the remaining 70\% assigned to the training set. For the subsequent R2 stage, aiming to mitigate potential biases introduced by the initial results and to prevent overfitting, an equal proportion of 20\% from both positive and negative examples is distributed to the validation set, another 20\% to the test set, with the remaining 60\% to the training set.

\subsection{Performance \& Ablation Evaluation}

In our study, we evaluate the performance of GasTrace's dual-round classification (R1 and R2), leveraging multidimensional metrics (i.e., precision, recall, F1 score, and overall accuracy). Moreover, all the testing results are derived from the test set. Furthermore, we document the processing times for both R1 and R2 phases to evaluate GasTrace's efficiency.

\begin{table}[htbp]
\caption{Comparison of Evaluation Indicators}
\label{tab: comparison}
\centering
\begin{tabular*}{\columnwidth}{@{\extracolsep{\fill}}lccccl}
\toprule
\textbf{Model} & \multicolumn{5}{c}{\textbf{Performance Metrics}} \\
\cmidrule(lr){2-6}
& \textbf{Precision} & \textbf{Recall} & \textbf{F1 Score} & \textbf{Accuracy} & \textbf{Time (s)} \\
\midrule
R1 & 0.9077 & 0.9035 & 0.9056 & 0.9259 & 1.1105 \\
R2 & 0.3733 & 0.5000 & 0.4275 & 0.7466 & 149.5622 \\
GasTrace & 0.9540 & 0.9603 & 0.9571 & 0.9673 & 172.9958 \\
\bottomrule
\end{tabular*}
\end{table}

As shown in Table~\ref{tab: comparison}, the classification precision at stage R1 is 90.77\%. However, the presence of a miss rate adversely affects other performance indicators, indicating that reliance solely on R1 is insufficient for accurately identifying malicious accounts. Table~\ref{tab: comparison} shows that the GasTrace with the R2 stage can alleviate this phenomenon, significantly improving the performance at several metrics (except for runtime). 

While skipping R1 and proceeding directly to the R2 stage, the performance is much weaker than R1. The reason is that lacking definition of network edge leads to the GAT model generally misclassifying accounts as normal.

\subsection{Generalization Capability Analysis}

We conduct a further in-depth comprehensive analysis, testing the generalization capability of GasTrace. 

\begin{figure}[h!]
  \centering
  \includegraphics[width=\linewidth]{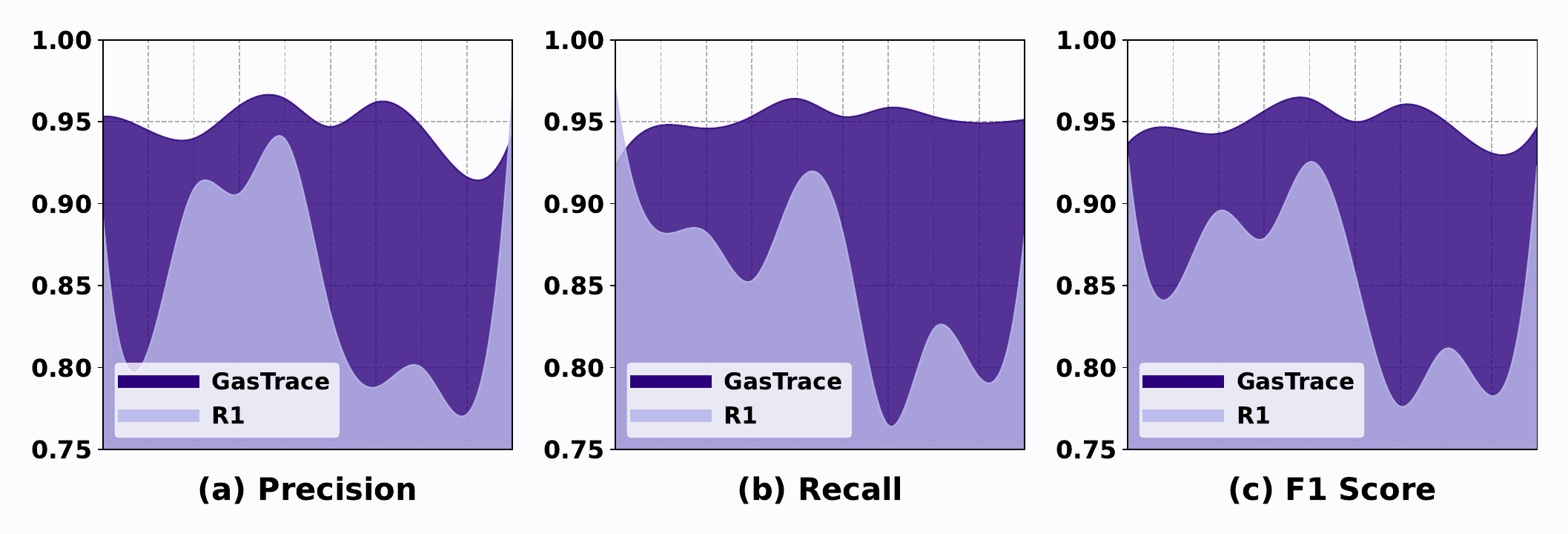}
  \caption{Area Chart: Precision, Recall, and F1 Score for R1, GasTrace.}
  \label{fig: area_chart}
\end{figure}

To more intuitively demonstrate the classification performance when executing R1 alone versus deploying the full GasTrace framework, we utilize area charts for a visual comparison of the two models in terms of precision, recall, and F1 score. It is important to note that these metrics will be normalized to eliminate the influence of the test set size.

As shown in the area chart in Figure~\ref{fig: area_chart}, the vertical dashed lines represent the ten tests conducted, with each dashed line intersecting the graph's edge curves indicating the performance of the two models on the respective evaluation metrics. Compared to R1 alone, GasTrace demonstrates a clear superiority in overall performance. Moreover, the smoother curve of GasTrace indicates its superior generalization capability.

\section{CONCLUSION}
In this study, we introduce GasTrace, a framework for detecting malicious accounts involved in sandwich attacks. We utilize a cascade classification strategy to focus on the gas features. The R1 stage extracts key features as the node features, constructing a node network graph. With the category prediction probabilities in R1, we complete the edge connections of the network graph. The R2 stage then makes a classification of the graph to identify malicious accounts. In the experimental phase, we demonstrate the performance of GasTrace, with a 96.73\% accuracy and a 95.71\% F1 score. Furthermore, we exhibit the generalization capability of GasTrace, which fluctuates less across all metrics.


\section{ACKNOWLEDGMENTS}

This work is sponsored by the National Natural Science Foundation of China (No.62362021), CCF-Tencent Rhino-Bird Open Research Fund (No.RAGR20230115), and Hainan Provincial Department of Education Project (No.HNJG2023-10).

\balance
\bibliographystyle{unsrt}
\bibliography{references}

\end{document}